# Progress in quantum electrodynamics theory of highly charged ions

*A. V. Volotka*[1,2,*], *D. A. Glazov*[2], *G. Plunien*[1], *and V. M. Shabaev*[2]

Recent progress in quantum electrodynamics (QED) calculations of highly charged ions is reviewed. The theoretical predictions for the binding energies, the hyperfine splittings, and the *g* factors are presented and compared with available experimental data. Special attention is paid to tests of bound-state QED at strong field regime. Future prospects for tests of QED at the strongest electric and magnetic fields as well as for determination of the fine structure constant and the nuclear magnetic moments with heavy ions are discussed.

## 1 Introduction

Quantum electrodynamics, being the relativistic quantum field theory of the electromagnetic force, describes all phenomena associated with electrically charged particles. Despite the mathematical complexity and difficulties caused by the occurrence of infrared and ultraviolet divergences, it has a great success in describing and predicting experimental results. For a long period of time quantum electrodynamics was mainly tested with light atomic systems: hydrogen, helium, positronium, and muonium. In these systems the QED effects are evaluated employing the expansion in two small parameters $\alpha$ and $\alpha Z$ ($\alpha$ is the fine structure constant and $Z$ is the nuclear charge number) and, therefore, are tested to the leading order(s) in these parameters.

Another scenario for tests of QED has appeared in experiments with highly charged ions. Heavy few-electron ions provide unique micro-laboratories for probing QED effects in the strongest electromagnetic fields accessible at present for experimental study [1, 2]. For example, at the surface of a uranium nucleus the electric field strength amounts to $|E| \simeq 2 \times 10^{19}$ V/cm, which is six orders of magnitude higher than the maximum electric field strength in a petawatt laser pulse. The magnetic field strength of the $^{209}$Bi$^{83}$ nucleus magnetic moment at the nuclear surface is about $10^9$ T, that is several orders of magnitude higher than the field of the most powerful magnets. In this regime, high-precision QED calculations become more complicated, since the consideration should be primarily relativistic. In particular, it means that the parameter $\alpha Z$ can not be utilized as an expansion parameter and, therefore, the calculations must be performed to all orders in $\alpha Z$. This requires developments of nonperturbative QED methods, which are suitable for the description of highly charged ions.

In this paper we review the current status of the QED calculations of the spectroscopic properties of highly charged ions: energy levels, hyperfine splitting, and *g* factor values. The relativistic units $\hbar = c = m_e = 1$ are used throughout the paper.

## 2 Binding energy

A systematic description of highly charged ions in the framework of QED starts with the one-electron Dirac equation

$$\left[-i\boldsymbol{\alpha}\cdot\boldsymbol{\nabla} + \beta + V(\mathbf{r})\right]\psi(\mathbf{r}) = E\psi(\mathbf{r}), \qquad (1)$$

where $V(\mathbf{r})$ is assumed to be the potential of the nucleus. Another choice of $V(\mathbf{r})$ is an effective local potential, which contains, besides the interaction with the nucleus, an approximate treatment of the interelectronic interaction. Solving the Dirac equation (1), one takes into account the interaction of the electron with the Coulomb field of the nucleus to all orders in $\alpha Z$. The interaction between the electron-positron and electromagnetic fields, which leads to the radiative and interelectronic-interaction corrections, is treated by the QED perturbation theory. The formulation of QED, in which the nucleus is treated as a classical source of the Coulomb field,

* Corresponding author  E-mail: andrey.volotka@tu-dresden.de
[1] Institut für Theoretische Physik, Technische Universität Dresden, Mommsenstraße 13, D-01062 Dresden, Germany
[2] Department of Physics, St. Petersburg State University, Oulianovskaya 1, Petrodvorets, 198504 St. Petersburg, Russia

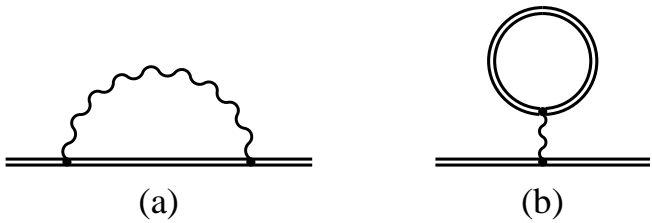

**Figure 1** Feynman diagrams representing the self-energy (a) and vacuum-polarization (b) radiative corrections. The wavy line indicates the photon propagator and the double line denotes electron propagating in the Coulomb field.

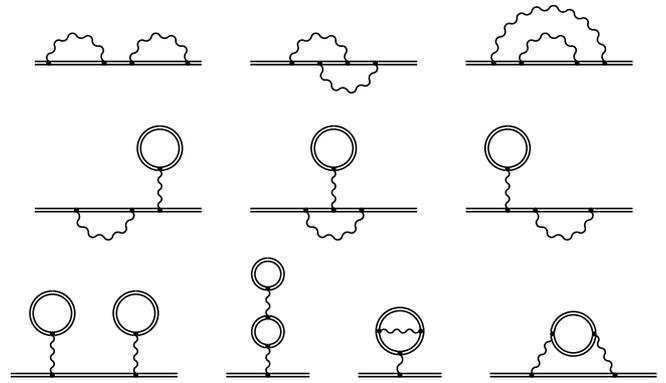

**Figure 2** Feynman diagrams representing the second-order one-electron radiative corrections.

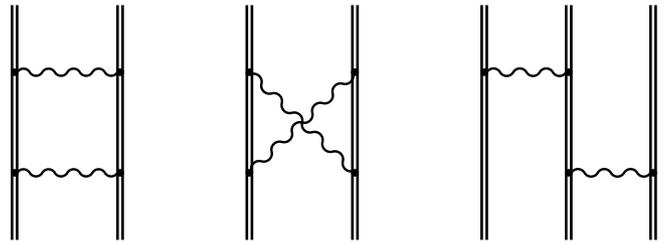

**Figure 3** Feynman diagrams representing the second-order interelectronic-interaction corrections.

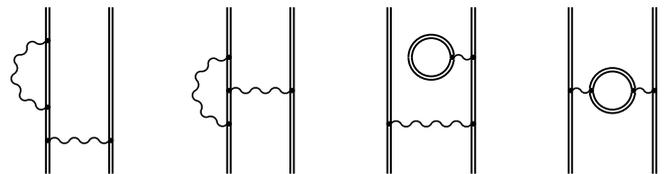

**Figure 4** Feynman diagrams representing the self-energy and vacuum-polarization screening corrections.

is known as the Furry picture of quantum electrodynamics.

For a point-like nucleus, the solution of the Dirac equation is known analytically, while for a finite-size nucleus this can be done either numerically (see, e.g., Ref. [3]) or analytically (Ref. [4]). The difference between the energies for the extended and the point-nucleus model is known as the finite-nuclear-size correction.

The radiative corrections of the first order in $\alpha$ are described by the Feynman diagrams depicted in Fig. 1. These are so-called self-energy (a) and vacuum-polarization (b) diagrams. In these diagrams, the double solid line indicates electron propagating in the Coulomb field of the nucleus, while the wavy line corresponds to a virtual photon. In contrast to light atomic systems, in highly charged ions these diagrams have to be calculated to all orders in $\alpha Z$. The nonperturbative evaluation of the self-energy correction was first performed by Desiderio and Johnson [5] employing the method proposed by Brown, Langer, and Schaefer [6]. Later, Mohr [7] developed a much more accurate and powerful method, which allowed him to carry out a high-precision evaluation of this correction in a wide range of $Z$ values. The most accurate calculations of the self-energy to all orders in $\alpha Z$ were performed in Refs. [8–11] for the point-like nucleus case and in Refs. [12, 13] for the extended nucleus case. The first nonperturbative calculations of the vacuum-polarization diagram was made by Soff and Mohr [14] and by Manakov, Nekipelov, and Fainshtein [15]. The most accurate results were obtained in Refs. [16, 17]. The second-order (two-loop) one-electron radiative corrections are defined by the diagrams depicted in Fig. 2. The complete nonperturbative calculations of these diagrams represent an extremely difficult task. Recent crucial developments in this respect were made in Refs. [18–20], where the complete set of the two-loop self-energy diagrams (the first three diagrams in Fig. 2) were rigorously evaluated. At present, only the last two diagrams in Fig. 2, being known only in the lowest order in $\alpha Z$ (see Ref. [21] and references therein), remain uncalculated to all orders in $\alpha Z$.

For few-electron ions, besides the one-electron radiative corrections, one has to take into account the interelectronic-interaction corrections. These corrections are suppressed by the parameter $1/Z$. For high-$Z$ ions this parameter becomes comparable with the fine structure constant $\alpha$, which characterizes the radiative corrections. The unperturbed many-electron wave functions are constructed within the $jj$-coupling scheme. The $jj$-coupling states are the eigenstates of the relativistic Hamiltonian of noninteracting electrons, which is the sum of one-electron Dirac Hamiltonians (1). For high-$Z$ regime, it is natural to employ the $jj$ coupling instead of the Russell-Saunders or the $LS$-coupling scheme, which become exact in the nonrelativistic limit. The calculations of the first-order interelectronic-interaction contributions are rather simple, while the second-order diagrams depicted in Fig. 3 are much more complicated. The first problem, which occurs in the treatment of these corrections, is the derivation of the formal expressions that are required for their numerical calculations. The most elaborate approach, which enables a rather simple derivation of the desired expressions and is applicable not only for a single state but also for degenerate and quasidegenerate, is the two-time Green's function method. This method was developed in Refs. [22, 23] and described in details in Ref. [24]. The complete QED calculation of the second-order (two-photon) exchange diagrams for the ground state of He-like ions was first performed by Blundell *et al.* [25] and by Lindgren *et al.* [26]. In the QED formulation the exact photon propagators are employed, that allows one to perform the calculations to all orders in $\alpha Z$ and obtain the gauge invariant results in each order of the perturbation theory. In the Breit approximation, which is frequently used in the many-body perturbation theory (MBPT) or configuration-interaction (CI) calculations, the photon propagator is treated approximately. This makes the Breit approximation valid only up to few lowest orders in $\alpha Z$. As a result, the MBPT calculations of the two-photon exchange corrections to the energy levels give the exact values only for the $\alpha^2$ and $\alpha^2(\alpha Z)^2$ terms. The contributions beyond the Breit approximation are referred to as the many-electron QED terms. Other many-electron QED contributions come from the combination of the radiative and interelectronic-interaction parts. They are known as the screened self-energy and screened vacuum-polarization contributions and depicted in Fig. 4. The rigorous calculations of the many-electron QED diagrams were performed in Refs. [27–30] for He-like ions and in Refs. [31–38] for Li-like ions.

So far we considered the Furry picture, where the nucleus is assumed to be a source of the external Coulomb field. Beyond this approximation one has to account for the finite nucleus mass and the intrinsic nuclear dynamics, that lead to the nuclear recoil and nuclear polarization effects. In contrast to the nonrelativistic theory, where the recoil effect for a hydrogenlike atom can easily be taken into account by using the electron reduced mass, the full relativistic theory of the recoil effect can be formulated only in the framework of QED. The complete relativistic formula for the recoil effect to first order in $m_e/M$ ($M$ is the nucleus mass) and to all orders in $\alpha Z$ was first derived in Ref. [39] (see also Ref. [40] and references therein) and numerically evaluated in Ref. [41].

The contributions discussed above can be precisely calculated order by order. However, this is not the case for the nuclear polarization corrections, which, due to the phenomenological description of the nucleon-nucleon interaction, set the ultimate accuracy limit up to which the QED corrections can be tested in highly charged ions. The energy shift due to this effect was evaluated by Plunien *et al.* [42, 43] and by Nefiodov *et al.* [44].

Finally, let us turn to the comparison with the experimental results. A precision of about $10^{-2}$ was obtained in the measurement of the ground state Lamb shift in the one-electron uranium ion [45]. This provides a test of QED effects on the level of about 2%. The most accurate measurements of the binding energy in highly charged heavy ions were performed with Li-like ions [46–48]. The $2p_{1/2} - 2s$ transition energy in $^{238}$U$^{89+}$ was measured to be 280.645(15) eV [48]. The total theoretical value for this transition energy, 280.71(10) eV [49], agrees well with the experimental result. Comparing the first- and second-order QED contributions with the total theoretical uncertainty, we find that the present status of the theory and experiment for Li-like uranium provides a test of QED on a 0.2% level to first order in $\alpha$ and on a 6% level to second order in $\alpha$.

## 3 Hyperfine structure

In case of a nonzero nuclear spin $I$, the atomic electron interacts also with the magnetic field induced by the nuclear magnetic moment $\mu = g_I I \mu_N$. Here $g_I$ is the nuclear $g$ factor and $\mu_N$ is the nuclear magneton. This interaction splits the energy levels into the hyperfine structure sublevels which correspond to different values of the total angular momentum of the ion $\mathbf{F} = \mathbf{I} + \mathbf{J}$, where $J$ is the total angular momentum of the electrons. Investigations of the hyperfine structure in highly charged ions are of particular interest since the electrons experience not only the strong electric Coulomb field but also the strong magnetic field. This provides a unique possibility for tests of QED in the strongest electric and magnetic fields.

Accurate measurements of the ground-state hyperfine splitting in heavy H-like ions: $^{209}$Bi, $^{165}$Ho, $^{185}$Re, $^{187}$Re, $^{207}$Pb, $^{203}$Tl, and $^{205}$Tl [50–54] have triggered a great interest in the theory of this effect. The ground-state hyperfine splitting in H-like ions can be represented

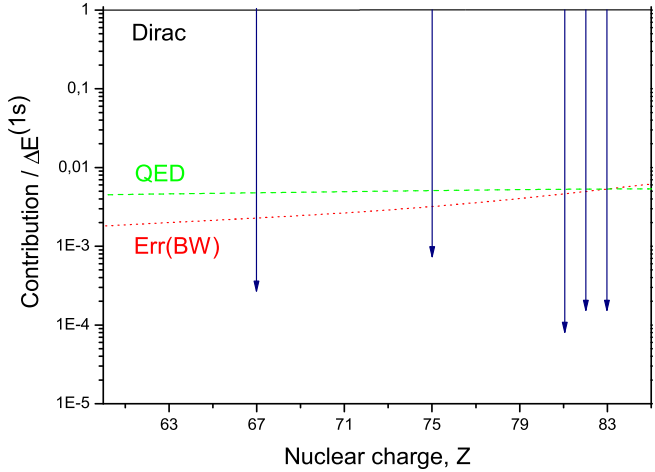

**Figure 5** (online color at: www.ann-phys.org) The relative contributions to the ground-state hyperfine splitting in H-like ions: the Dirac value, the QED correction, and the uncertainty of the BW effect. The vertical lines with arrows represent the accuracy of the existing experimental results [50–54].

in the form:
$$\Delta E^{(1s)} = \Delta E^{(1s)}_{\text{Dirac}}(1 - \epsilon^{(1s)}) + \Delta E^{(1s)}_{\text{QED}}, \qquad (2)$$

where $\Delta E^{(1s)}_{\text{Dirac}}$ is the relativistic (Dirac) value of the 1s hyperfine splitting, which also contains the nuclear charge distribution correction, $\epsilon^{(1s)}$ represents the nuclear magnetization distribution correction, so called Bohr-Weisskopf (BW) effect, and $\Delta E^{(1s)}_{\text{QED}}$ stands for the radiative correction. The radiative correction was evaluated independently by several groups [55–62] and the results are found in a good agreement with each other. The theoretical uncertainty is mainly determined by the BW effect which is very sensitive to the nuclear model employed in the calculation. In Fig. 5 the relative contributions of the Dirac value, the QED correction, and the uncertainty of the BW effect evaluated within the single-particle nuclear model [58] are presented. As one can see from the figure the uncertainty of the nuclear magnetization distribution correction strongly masks the QED contribution. Accordingly, the direct identification of the QED effects on the hyperfine splitting in heavy H-like ions appeared to be unfeasible.

In this context, it was proposed to consider a specific difference of the ground state hyperfine splitting values in H- and Li-like ions [63]:
$$\Delta' E = \Delta E^{(2s)} - \xi \Delta E^{(1s)}, \qquad (3)$$

where the parameter $\xi$ is chosen to cancel the Bohr-Weisskopf correction. The parameter $\xi$ can be calcu- lated to a rather high accuracy, because it is determined mainly by the behavior of the electron wave function at the atomic scale and, therefore, almost independent of the nuclear structure. The ground-state hyperfine splitting in Li-like ions $\Delta E^{(2s)}$ is conveniently written in the form:

$$\Delta E^{(2s)} = \Delta E^{(2s)}_{\text{Dirac}}(1 - \epsilon^{(2s)}) + \Delta E^{(2s)}_{\text{QED}}$$
$$+ \Delta E_{\text{int}}(1 - \epsilon^{(\text{int})}) + \Delta E_{\text{SQED}}. \qquad (4)$$

Here $\Delta E^{(2s)}_{\text{Dirac}}$ is the one-electron relativistic value of the 2s hyperfine splitting, $\epsilon^{(2s)}$ and $\epsilon^{(\text{int})}$ denote the BW corrections to the leading and the interelectronic-interaction terms, respectively, $\Delta E^{(2s)}_{\text{QED}}$, $\Delta E_{\text{int}}$, and $\Delta E_{\text{SQED}}$ represent the one-electron QED, the interelectronic-interaction, and the screened QED corrections. In Fig. 6 we display the $Z$-dependence of the corresponding contributions for the hyperfine splitting in Li-like ions. According to this figure, like in case of H-like ions, the uncertainty of the BW correction masks the QED contributions. But this uncertainty can be substantially reduced in $\Delta' E$ defined by Eq. (3). The relative contributions of the individual terms to the specific difference $\Delta' E$ are presented in Fig. 7. As one can see from the figure, the remaining uncertainty of the BW effect is two orders of magnitude smaller than the screened QED or two-photon exchange corrections. Thereby, the stringent tests of QED in combination of the strong electric and magnetic fields can be achieved by studying the specific difference of the hyperfine splitting values in H- and Li-like ions.

As was mentioned above, to date there exist several accurate measurements of the hyperfine splitting in H-like ions. But this is not the case for Li-like ions. The first measurement of the hyperfine splitting in Li-like $^{209}$Bi was made in Ref. [64]. However, since this was an indirect measurement, its uncertainty is rather large. After more than a decade of search, this transition line has been observed in a laser spectroscopy experiment at GSI [65]. It is expected that at the HITRAP facility at GSI the experimental accuracy will be improved by several orders of magnitude approaching the relative uncertainty several parts in $10^{-7}$ [66, 67].

Achievement of the required theoretical accuracy for the specific hyperfine splitting difference for H- and Li-like heavy ions demands the rigorous evaluation of various QED and interelectronic-interaction effects. Since the influence of the one-electron QED corrections is considerably reduced in the difference, the total value of $\Delta' E$ is essentially determined by the screened QED and interelectronic-interaction corrections. These contributions correspond to the third-order terms in the QED perturbation theory expansion. The generic types of the

           

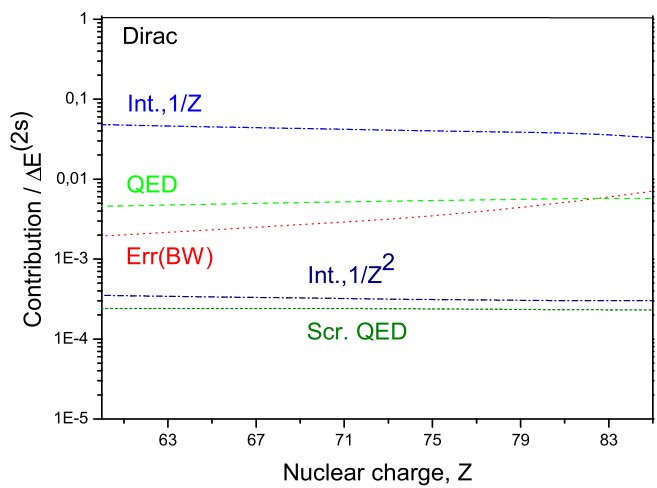

**Figure 6** (online color at: www.ann-phys.org) The relative contributions to the ground-state hyperfine splitting in Li-like ions: the Dirac value, the one-electron QED, interelectronic-interaction, and screened QED corrections, together with the uncertainty of the BW effect.

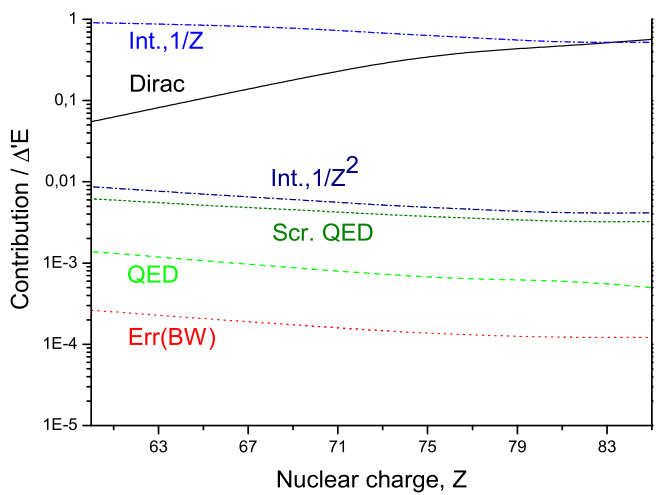

**Figure 7** (online color at: www.ann-phys.org) The relative contributions to the specific difference of the ground state hyperfine splitting values in H- and Li-like ions: the Dirac value, the one-electron QED, interelectronic-interaction, and screened QED corrections, together with the remaining uncertainty of the BW effect.

screened self-energy and vacuum-polarization diagrams are presented in Figs. 8 and 9. Each diagram contains three parts: self-energy or vacuum-polarization loops, interelectronic interaction, and the vertex with an addi-tional magnetic potential. As an external potential we employ the hyperfine interaction or the Zeeman interaction potentials. Taking into account the permutations of the one-electron states in these diagrams, one obtains 36 screened self-energy and 32 screened vacuum-polarization contributions, respectively. The screened self-energy contribution has been calculated rigorously within the systematic QED approach in Refs. [68, 69]. This calculation represents an essential advance beyond the local screening potential approximation employed in the previous works [62, 70–74]. The screened vacuum-polarization contribution has been evaluated in the free-loop approximation in Refs. [68, 69], and later the major part of the diagrams has been calculated to all orders in $\alpha Z$ [75]. Presently, only the last two diagrams in Fig. 9 remain uncalculated to all orders in $\alpha Z$.

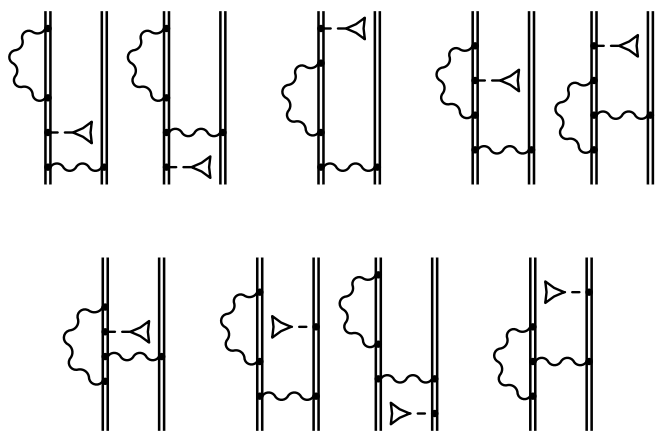

**Figure 8** Feynman diagrams representing the screened self-energy corrections in the presence of an external potential. The dashed line terminated with the triangle denotes the interaction with the magnetic field.

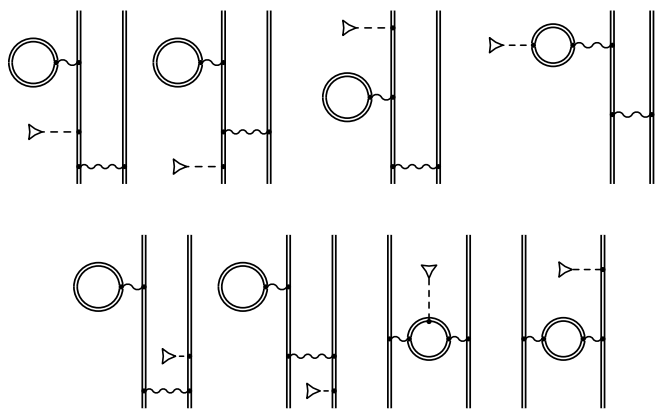

**Figure 9** Feynman diagrams representing the screened vacuum-polarization corrections in the presence of an external potential.

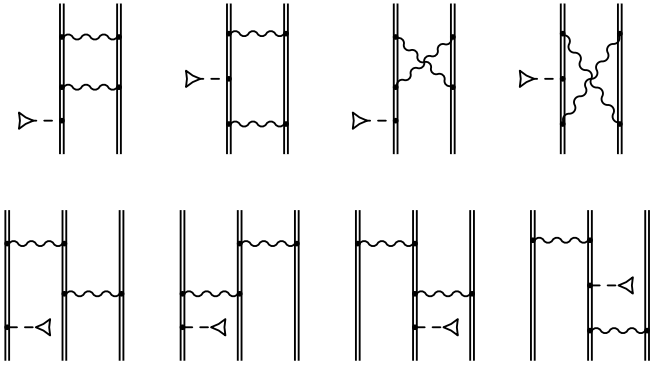

**Figure 10** Feynman diagrams representing the two-photon exchange corrections in the presence of an external potential.

**Table 1** Individual contributions to the specific difference $\Delta'E$ for $^{209}$Bi in meV.

| Effect | $\xi\Delta E^{(1s)}$ | $\Delta E^{(2s)}$ | $\Delta'E$ |
|---|---|---|---|
| Dirac value | 876.638 | 844.829 | −31.809 |
| QED | −5.088 | −5.052 | 0.036 |
| Interel. interaction | | | |
| $\sim 1/Z$ | | −29.995 | −29.995 |
| $\sim 1/Z^2$ | | 0.258 | 0.258 |
| $\sim 1/Z^{3+}$ | | −0.003(3) | −0.003(3) |
| Screened QED | | 0.193(2) | 0.193(2) |
| Total | | | −61.320(4)(5) |

The rigorous calculations of the interelectronic-interaction terms are performed within the $1/Z$ perturbation expansion. The one-photon exchange correction corresponding to the first-order in $1/Z$ was evaluated rigorously in Ref. [76]. Until recently, the rigorous calculation of the two-photon exchange diagrams (Fig. 10) remained a challenge for theory. In Refs. [62, 77, 78], the contributions of the second and higher orders in $1/Z$ have been calculated within the Breit approximation employing the MBPT and CI methods. Recently, the evaluation of the two-photon exchange diagrams has been performed rigorously to all orders in $\alpha Z$ in Ref. [79].

The current status of the specific difference of the hyperfine splitting values of H- and Li-like Bi is presented in Table 1. In this case, the cancellation of the BW effect appears at $\xi = 0.16886$, while the specific difference amounts to 61.320(4)(5) meV. The first uncertainty originates from the uncalculated parts of the screened vacuum-polarization contributions. The second uncertainty comes from the nuclear magnetic moment ($\mu = 4.1106(2)\mu_N$ [80]), the nuclear polarization corrections [81], and other nuclear effects, which are not completely canceled in the specific difference. Thus, the theoretical accuracy achieved for the specific difference allows one to test the many-electron QED effects at the level of a few percent, provided the hyperfine splittings in H- and Li-like bismuth are measured with a relative accuracy of about $10^{-6}$. When the QED corrections will be tested and found to be valid, the comparison between the theoretical and experimental values for H- and Li-like ions will enable the determination of the nuclear magnetic moments and their volume distribution.

## 4 *g* factor

In a homogeneous magnetic field the energy levels split according to the projection of the ion's angular momentum on the field direction. For a weak magnetic field strength $B$, such a splitting is known as the Zeeman splitting. For a spinless nucleus this splitting can be written in the form:

$$\Delta E(B) = g M_J \mu_B B, \quad (5)$$

where $\mu_B$ is the Bohr magneton, $M_J$ is the projection of the angular momentum on the field direction, and the *g* factor is a dimensionless quantity characterizing the energy shift.

In recent years, a spectacular progress was made in the experimental and theoretical investigations of the bound-electron *g* factor. High-precision measurements of the ground state *g* factor of H-like carbon [82] and oxygen [83] and the related theoretical calculations provided determination of the electron mass with an accuracy which is four times better than that of the previously accepted value. Recently, highly accurate measurements have been performed for the *g* factor of H-like $^{28}$Si$^{13+}$ [84, 85] with a statistical uncertainty significantly smaller than the uncertainty coming from the electron mass value. To date, these experiments provide the most stringent test of the one-electron QED corrections in the presence of a magnetic field. Accurate measurement of the *g* factor of Li-like $^{28}$Si$^{11+}$ has been recently accomplished in Ref. [86]. High-precision measurements are also anticipated for B-like $^{40}$Ar$^{13+}$ and $^{40}$Ca$^{15+}$ [87, 88]. The investigations of the *g* factor of few-electron ions provide an access to the many-electron QED corrections. Extentions of these investigations to high-$Z$ ions will provide a great opportunity to probe the magnetic sector of QED at a strong Coulomb field.

The ground-state $g$ factors of H- and Li-like ions are conveniently written in the form:

$$g^{(1s)} = g^{(1s)}_{\text{Dirac}} + \Delta g^{(1s)}_{\text{QED}} + \Delta g^{(1s)}_{\text{nucl}}, \tag{6}$$

$$g^{(2s)} = g^{(2s)}_{\text{Dirac}} + \Delta g^{(2s)}_{\text{QED}} + \Delta g_{\text{int}} + \Delta g_{\text{SQED}} + \Delta g^{(2s)}_{\text{nucl}}, \tag{7}$$

where $g^{(1s)}_{\text{Dirac}}$ and $g^{(2s)}_{\text{Dirac}}$ are the one-electron relativistic values of the $1s$ and $2s$ $g$ factors for the point-charge nucleus, $\Delta g^{(1s)}_{\text{QED}}$ and $\Delta g^{(2s)}_{\text{QED}}$ are the one-electron QED corrections, $\Delta g^{(1s)}_{\text{nucl}}$ and $\Delta g^{(2s)}_{\text{nucl}}$ incorporate the nuclear-size, nuclear-recoil and nuclear-polarization corrections, $\Delta g_{\text{int}}$ and $\Delta g_{\text{SQED}}$ denote the interelectronic-interaction and screened QED contributions, respectively. Evaluations of the values $g^{(1s)}_{\text{Dirac}}$ and $g^{(2s)}_{\text{Dirac}}$ cause no problem. The one-electron QED corrections are evaluated employing the perturbation theory in the parameter $\alpha$. The first-order QED corrections were addressed in Refs. [89–98], while the second-order corrections were evaluated within $\alpha Z$-expansion in Refs. [99–101]. The nuclear effects on the $g$ factor have been also investigated: the nuclear size correction was derived analytically in Ref. [102], the nuclear deformation correction was calculated in Ref. [103], the recoil contribution to first order in $m_e/M$ and to all orders in $\alpha Z$ was derived in Ref. [104] and numerically evaluated in Ref. [105], the nuclear polarization correction was investigated in Ref. [106]. For Li-like ions, besides the one-electron corrections, one has to take into account the screened radiative and interelectronic-interaction corrections, which are defined by diagrams similar to those for the hyperfine splitting (Figs. 8, 9, and 10). For low-$Z$ ions, the screened radiative corrections were obtained employing the perturbation theory to the leading orders in $\alpha Z$ in Refs. [94, 107]. For middle-$Z$ ions, the screening effect was evaluated by introducing the effective screening potential in the QED calculations to all orders in $\alpha Z$ [70]. For high-$Z$ ions, the most accurate results for the screened radiative corrections have been obtained rigorously within the systematic QED approach [68, 69]. The one-photon exchange diagrams, which represent the interelectronic-interaction corrections of the first order in $1/Z$, were evaluated in the framework of QED in Ref. [108]. Recently, the two-photon exchange diagrams have been rigorously evaluated for the case of Li-like $^{28}$Si$^{11+}$ [86]. The individual contributions to the $g$ factor of Li-like silicon are presented in Table 2. As one can see, the total theoretical value is in excellent agreement with the experimental one. It confirms the interelectronic-interaction effects at the level of $10^{-4}$ and, in particular, the two-photon exchange contribution is probed on a 1% level.

**Table 2** Individual contributions to the ground-state $g$ factor of Li-like $^{28}$Si$^{11+}$.

| Effect | $^{28}$Si$^{11+}$ |
|---|---|
| Dirac value | 1.998 254 751 |
| Finite nuclear size | 0.000 000 003 |
| QED, $\sim \alpha$ | 0.002 324 044 (3) |
| QED, $\sim \alpha^{2+}$ | −0.000 003 517 (1) |
| Interelectronic interaction, $\sim 1/Z$ | 0.000 321 592 |
| Interelectronic interaction, $\sim 1/Z^2$ | −0.000 006 876 (1) |
| Interelectronic interaction, $\sim 1/Z^{3+}$ | 0.000 000 085 (22) |
| Screened QED | −0.000 000 212 (46) |
| Nuclear recoil | 0.000 000 039 (1) |
| Total | 2.000 889 909 (51) |
| Experiment [86] | 2.000 889 889 9(21) |

Table 3 presents the individual contributions to the ground-state $g$ factors of H- and Li-like lead. Here we split the QED corrections into the free and bound-state QED parts. The free QED terms corresponding to the anomalous magnetic moment of the free electron are known through the order $\alpha^5$ [109] and do not depend on the nuclear charge $Z$. The bound-state QED terms reflect the binding effects on the QED corrections and rapidly increase with the nuclear charge. However, as one can see from Table 3, the theoretical uncertainties due to the nuclear size effect become comparable with the bound-state QED corrections of second order in $\alpha$. This strongly restricts the tests of bound-state QED in such investigations. In Ref. [108], it was shown that the uncertainty due to the nuclear effects can be significantly reduced in a specific difference of the $g$ factors of H- and Li-like ions with the same nucleus, similar to the difference of the hyperfine splitting values (see Eq. (3)). Therefore, studying this difference, the QED effects can be investigated to a much higher accuracy than in the separate investigations with H- or Li-like ions.

Besides a test of QED, investigations of the $g$ factors of highly charged ions can provide a possibility for an independent determination of the fine structure constant from the bound-state QED at the high-$Z$ regime [110]. For this purpose it was proposed to consider a specific difference of the $g$ factors of H- and B-like ions of the same spinless isotope in the lead region. It was found, that in case of lead this specific difference can be calcu-lated to an accuracy of about $10^{-10}$. Together with the corresponding experimental results for the $g$ factors of H- and B-like lead, this may lead to a determination of $\alpha$ to a precision comparable to one obtained from the free-electron $g$ factor [109].

**Table 3** Individual contributions to the ground-state $g$ factors of $^{208}$Pb$^{81+}$ and $^{208}$Pb$^{79+}$.

| Effect | $^{208}$Pb$^{81+}$ | $^{208}$Pb$^{79+}$ |
|---|---|---|
| Dirac value | 1.734 947 023 | 1.932 002 904 |
| Nuclear size | 0.000 452 9(8) | 0.000 078 58(13) |
| Free QED | | |
| $\sim \alpha$ | 0.002 322 819 | 0.002 322 819 |
| $\sim \alpha^2$ and higher orders | −0.000 003 515 | −0.000 003 515 |
| Bound-state QED | | |
| $\sim \alpha$ | 0.000 561 50(2) | 0.000 088 9(1) |
| $\sim \alpha^2$ and higher orders | −0.000 000 2(6) | −0.000 000 1(5) |
| Interelectronic interaction | | 0.002 140 7(27) |
| Screened QED | | −0.000 001 8(2) |
| Nuclear recoil | 0.000 001 723 | 0.000 000 25(35) |
| Nuclear polarization | −0.000 000 2(1) | −0.000 000 04(2) |
| Total | 1.738 282 0(10) | 1.936 628 7(28) |

For ions with nonzero nuclear spin $I$ the energy shift depends not only on the electronic $g$ factor but also on the nuclear $g$ factor $g_I = \mu/(\mu_N I)$. The energy level structure depends on the ratio between the Zeeman and the hyperfine splitting values. For weak magnetic fields, the Zeeman interaction can be treated perturbatively and the hyperfine structure sublevels split into the Zeeman patterns. If the Zeeman splitting is comparable to the hyperfine splitting, the energy level structure is described by the Breit-Rabi formula. Theoretical and experimental investigations of these splittings can provide determinations of the nuclear magnetic moments on the $10^{-6}$ accuracy level. The case of a weak magnetic field was investigated in Refs. [111–114], while the corrections to the Breit-Rabi formula were evaluated in Refs. [115, 116].

## 5 Conclusion

In this paper we have reviewed the present status of the QED calculations of highly charged ions. The comparison between the theoretical results and the corresponding experimental data shows that at present the best test of QED at strong electric fields has been achieved in the investigations of the binding energies. For the case of Li-like uranium the $2p_{1/2} - 2s$ transition energy provides a test of bound-state QED on a 0.2% accuracy level to first order in $\alpha$ and on a 6% accuracy level to second order in $\alpha$. The Bohr-Weisskopf effect restricts the direct identification of the QED effects on the hyperfine splitting in heavy H-like ions. It was shown instead that the theoretical uncertainty can be significantly reduced in a specific difference of the hyperfine splitting values of H- and Li-like ions with the same nucleus. Thus, the investigations of the hyperfine splitting in heavy H- and Li-like ions of the same isotope provide a unique opportunity for tests of the bound-state QED in combination of the strong electric and magnetic fields. The theoretical accuracy achieved for the specific difference between the hyperfine splittings values in H- and Li-like bismuth allows us to identify the many-electron QED effects at the level of a few percent. The present experimental and theoretical investigations of the bound-electron $g$ factor provide the stringent tests of the magnetic sector of bound-state QED. The one-electron QED corrections to the bound-electron $g$ factor have been probed by direct measurements with H-like carbon, oxygen, and silicon, while the measurement of the $g$ factor of Li-like silicon yields the most stringent test of the many-electron QED effects in presence of a magnetic field. Extentions of these measurements to high-$Z$ ions and to ions with nonzero nuclear spin can serve for independent determinations of the fine structure constant and the nuclear magnetic moments.

**Acknowledgements.** The investigations reported in this paper were supported by the DFG (Grant No. VO 1707/1-2), GSI, RFBR (Grants No. 12-02-31803 and 13-02-00630), the Ministry of Education and Science of Russian Federation (Grant No. 8420). D.A.G. acknowledges financial support by the FAIR – Russia Research Center and by the "Dynasty" foundation.

**Key words.** Quantum electrodynamics, highly charged ions, binding energy, hyperfine structure, $g$ factor.